# Incorporating Disciplinary Practices Into Characterizations of Progress in Responsive Teaching

Jennifer Richards and Andrew Elby
Department of Teaching and Learning, Policy and Leadership, University of Maryland, 2311 Benjamin Building, College Park, MD 20742

Ayush Gupta
Department of Physics, University of Maryland, 1320 Physics Building, College Park, MD 20742

Author Note

Jennifer Richards, the corresponding author, is now at Department of Curriculum & Instruction, University of Washington, Box 353600, Seattle, WA 98195. Phone 443-794-3192. Email jrich14@uw.edu.

A preliminary version of this article was presented at ICLS 2014. This work was supported by funding from NSF DRL-0733613 and NSF EHR/DUE-0831970. All findings, opinions, and recommendations expressed herein are those of the authors and do not necessarily reflect the views of the National Science Foundation. Special thanks to David Hammer, Paul Hutchison, Eric Kuo, and Amy Robertson for feedback on earlier versions of the manuscript. The manuscript is currently in the review process.

Abstract

Responsive teaching, in which teachers adapt instruction based on close attention to the substance of students' ideas, is typically characterized along two dimensions: the level of detail at which teachers attend and respond to students' ideas, and the stance teachers take toward what they hear – evaluating for correctness vs. interpreting meaning. We propose that characterizations of progress in responsive teaching should also consider the disciplinary centrality of the practices teachers notice and respond to within student thinking. To illustrate what this kind of progress can look like, we present a case study of a middle school science teacher who implemented the "same" lesson on the motion of freely falling objects in two subsequent years. We argue that his primary shift in responsiveness stemmed from a shift in which disciplinary practices he preferentially noticed and foregrounded. He moved from a focus on causal factors or variables to a more scientifically productive focus on causal stories or explanations. We explore how participation in a professional development community, institutional constraints, and a shift in personal epistemology may have contributed to the nature and stability of this shift in responsiveness.

Incorporating Disciplinary Practices Into Characterizations of Progress in Responsive Teaching

In the mathematics and science teacher education and professional development literature, a consensus is emerging about the need for teachers to elicit, attend to, and make instructional adjustments in response to the substance (as opposed to just the correctness or incorrectness) of students' ideas (e.g., Brodie, 2011; Coffey, Hammer, Levin, & Grant, 2011; Gallas, 1995; Hammer & van Zee, 2006; Kazemi & Hintz, 2014; Lampert et al., 2013; Lineback, 2014; NCTM, 2000; NRC, 2012; Rosebery & Warren, 1998; Sherin, Jacobs, & Philipp, 2011; Windschitl, Thompson, Braaten, & Stroupe, 2012). These practices of noticing/attending/responding to student thinking – which we will refer to under the umbrella term "responsive teaching" (Hammer, Goldberg, & Fargason, 2012, p. 54) – are central to developing programs of study on ambitious teaching (e.g., Lampert et al., 2013; Windschitl et al., 2012) and correlate with deeper student learning (e.g., Carpenter, Fennema, Peterson, Chiang, & Loef, 1989; Pierson, 2008). Partly for these reasons, researchers have started to operationalize conceptualizations of what counts as more responsive vs. less responsive teaching, in order to clarify the underlying construct(s) and to enable assessment of whether an intervention succeeds in fostering greater responsiveness among teachers. As we demonstrate below, within the responsive teaching literature, improvement along one or both of the following two dimensions typically counts as progress toward more responsive teaching: attending to a greater *level of detail* in students' ideas, and taking a more interpretive as opposed to evaluative *stance* toward those ideas.

In this paper, we propose an additional notion of progress in responsive teaching: attending to more central disciplinary practices within the substance of student thinking. To be clear, other pockets of literature about teaching practice and classroom culture emphasize the teacher's role of hearing multiple seeds of disciplinary practice in student thinking and responding in ways that promote particular practices and norms (e.g., Ball, 1993; Chazan & Schnepp, 2002; Cobb, Yackel, and Wood, 1989; Hammer, 1997). Our aim is to bring these considerations into the conversation on responsive teaching, pressing for articulation of the various aspects of students' ideas teachers attend and respond to (beyond correctness) and how different foci may reflect and promote different views of what counts as authentic disciplinary activity.

Our argument will flow as follows. First, we review current conceptualizations of responsive teaching, establishing that these conceptualizations do not currently attend to the extent to which a teacher notices the beginnings of different disciplinary practices in student thinking and gears responses to nurturing the most central or productive practices (given the task at hand and the emerging flow of discussion). We then briefly review examples from scholars who explicitly grapple with and highlight specific disciplinary aspects of student thinking before turning to the case study at the center of our paper. The case of a middle school science teacher who taught a similar lesson in two different years showcases how differences in responsive teaching may not primarily involve differing levels of specificity or stance, but rather differences in what the teacher foregrounds in his attention and responsiveness to student thinking. In year 1, the teacher foregrounded the *causal factors* offered by students – variables that students thought would affect the motion of the object under consideration. In year 2, by contrast, the teacher foregrounded students' *causal stories*, their explanations of *why* given factor(s) would produce the motion predicted. We argue that, in the context of these this lesson, causal storytelling is more disciplinarily productive than (merely) identifying causal factors, and as such the teacher demonstrated progress in responsive teaching – progress that is grounded in the nature of the disciplinary practices to which he attended and responded. Finally, we turn attention to two questions raised by our case study:
> (i) Why did our focal teacher exhibit a different—and we would argue, better—pattern of attention and responses in year 2?
> (ii) How can researchers distinguish between stable progress in responsive teaching and context-dependent "fluctuations?"

In making this argument and exploring these issues, we hope to spark discussion of how considerations of disciplinary practices can be incorporated into notions of responsive teaching.

## Current characterizations of responsive teaching

Researchers generally characterize responsive teaching along two interrelated dimensions: the *level of detail* at which teachers attend and respond to students' ideas (e.g., Crespo, 2000; Jacobs, Lamb, & Philipp, 2010), and the *stance* (evaluating vs. interpreting students' ideas) teachers take toward what they hear (e.g., Brodie, 2011; Levin, Hammer, & Coffey, 2009). In what follows, we describe each dimension and illustrate how the literature instantiates both. We then examine more closely two multi-leveled schemes for characterizing greater vs. lesser responsiveness – video club work from Sherin and van Es (e.g., Sherin & van Es, 2009; van Es, 2011) and the Cognitively Guided Instruction (CGI) project (e.g., Fennema et al., 1996; Franke & Kazemi, 2001) – chosen because of their careful documentation and their prominence in the literature. In doing so, we demonstrate how distinctions between the levels incorporate the *level of detail* and *stance* dimensions. The video club work, like most in the literature, focuses on all aspects of students' ideas within the given domain (in this case mathematics) (van Es, 2011). CGI focuses on a specific facet of student thinking in particular subdomains of elementary school mathematics, such as students' problem-solving strategies for addition and subtraction (Franke & Kazemi, 2001). Despite these differing foci, both schemes define progress in largely the same way, as movement from superficial attention to students' ideas or strategies to more detailed attention and interpretation of what students are doing.

**Dimension 1: Level of detail**

A common consideration in characterizing and evaluating responsiveness is whether teachers' descriptions of students' reasoning are (a) general, drawing on superficial aspects of an individual's ideas or on how "the class" is reasoning; or (b) specific, drawing on details and nuances within the ideas of individual students, with the latter considered more responsive (Crespo, 2000; Fennema et al., 1996; Franke, Carpenter, Fennema, Ansell, & Behrend, 1998; Franke, Carpenter, Levi, & Fennema, 2001; Jacobs, Lamb, & Philipp, 2010; Jacobs, Lamb, Philipp, & Schappelle, 2011; Kazemi & Franke, 2004; Levin & Richards, 2011; Lineback, 2014; Sherin & Han, 2004; Sherin & van Es, 2009; van es, 2011; van Es & Sherin, 2008, 2010). For instance, in a work group in which teachers were expected to share how their students approached a particular mathematics problem, Kazemi and Franke (2004) noted that early on, teachers paid little attention to the specifics of students' solutions. They focused largely on whether students' strategies were correct (an evaluative stance as discussed below). Similarly, Crespo (2000) noted that the pre-service teachers in her study initially made claims about student understanding that were not grounded in much evidence; but later, teachers' "comments revealed greater attention towards the meaning of student's mathematical thinking rather than surface features" (p. 170). In these examples, progress in attending and responding to student thinking resided in the specificity of attention the teachers demonstrated with respect to students' ideas.

**Dimension 2: Stance toward ideas**

Another way in which researchers characterize favorable shifts in attending and responding to student thinking is in movement from an evaluative stance, looking only for correctness, to an interpretive stance, making sense of students' ideas (e.g., Brodie, 2011; Crespo, 2000; Davis, 1997; Empson & Jacobs, 2008; Goldsmith & Seago, 2011; Levin, Hammer, & Coffey, 2009; Levin & Richards, 2011; Pierson, 2008; Sherin & van Es, 2009; van Es, 2011; van Es & Sherin, 2008, 2010). For instance, Goldsmith and Seago (2011) described how, during early meetings of mathematics teacher work groups, the teachers interpreted student work in light of the correct answer. In later meetings they looked for the logic in students' solutions. van Es and Sherin's research on video clubs includes *stance* as an explicit dimension along which they consider teachers' contributions, noting whether teachers merely describe students' ideas, evaluate them, or seek to interpret their meaning.

Studying teachers' moves in classroom discourse, researchers infer *stance* from teachers' follow-up moves. For example, Empson and Jacobs (2008) define a progression in listening expertise that moves from (i) directive listening, where the teacher focuses on alignment between a student's idea and an expected response and actively seeks to elicit the expected response, to (ii) observational listening, where the teacher passively listens to students' ideas without seeking to extend them, to (iii) responsive listening, where the teacher actively probes students' ideas and seeks to understand and build on the details. Similarly, Brodie (2011) and Pierson (2008) differentiate between follow-up moves that are corrective versus interpretive in nature, such as trying to elicit particular (often canonically correct) ideas from students versus pressing for more information about what students are offering. In these examples, a shift toward greater responsiveness is conceptualized as a change from seeking to evaluate students' ideas to seeking to understand them in more depth.

**A closer look at multi-level schemes**

Here, we take a closer look at two multi-level schemes in the literature that define and describe progress in noticing and/or responding to students' ideas – video club work from van Es and Sherin, and the CGI project. We selected these two schemes because they are both well-articulated, with clear descriptions and examples of each level, and well-documented across several papers from each project. In addition, they are highly cited in the literature. In what follows, we demonstrate how the levels in each scheme closely correspond to the dimensions identified above.

**Noticing in the context of video clubs.** The first professional development effort we describe is Sherin and van Es's work with elementary- and middle-school math teachers in video clubs (Sherin & van Es, 2009; van Es, 2011; van Es & Sherin, 2008, 2010; building on earlier work from Sherin & Han, 2004). Video clubs bring groups of teachers together to watch and reflect on video of classroom instruction. These specific video clubs aimed to help teachers learn to notice and interpret students' mathematical thinking during the flow of instruction. To this end, the facilitators "prompted the teachers to examine students' ideas about the mathematics, to use evidence to support claims they made about students' thinking, and to interpret the students' understanding about the mathematics" (van Es & Sherin, 2008, p. 248). Note that the clubs did not explicitly involve discussion of potential responses or instructional moves, though these came up spontaneously (van Es, 2011).

van Es and Sherin (2008) developed codes to capture important aspects of teachers' video club contributions. The first dimension is Actor, whether the video club participant discusses a teacher, student, or other actor in the video. The second dimension is Topic (e.g., mathematical thinking, management, etc.). The third and fourth dimensions correspond to the ones we reviewed above -- Stance (describe, evaluate, or interpret) and Specificity (general or specific). These last two dimensions also constituted a large part of van Es's (2011) tiered framework for learning to notice student thinking in the video club context. For instance, part of what distinguishes Level 1 (baseline noticing) from Level 3 (focused noticing) is the specificity or level of detail at which video club participants attend to students' ideas. Teachers at Level 1 "form general impressions of what occurred" (p. 139) and "provide little or no evidence to support analysis" (p. 139), whereas teachers at Level 3 "attend to particular students' mathematical thinking" (p. 139) and "refer to specific events and interactions as evidence" (p. 139). The stance teachers take toward the ideas they hear is also integrated into the framework; teachers at Level 2 (mixed noticing) "provide primarily evaluative with some interpretive comments" (p. 139), whereas teachers at higher levels shift to entirely interpretive comments. In other words, more sophisticated video club participation entailed a focus on interpreting students' specific mathematical ideas.

**Cognitively Guided Instruction.** Another effort to help teachers base their instruction around the substance of students' mathematical thinking is the CGI project (Carpenter, Fennema, & Franke, 1996; Carpenter, Fennema, Franke, Levi, & Empson, 2000; Carpenter, Fennema, Peterson, Chiang, & Loef, 1989; Fennema et al., 1996; Fennema, Franke, Carpenter, & Carey, 1993; Franke et al., 1998, 2001; Franke & Kazemi, 2001). Drawing on research on children's conceptual development in specific subdomains of mathematics such as addition and subtraction, researchers developed principled frameworks outlining problem categories in these content areas and strategies children often use to solve problems of that type (Franke & Kazemi, 2001). They anticipated that familiarizing teachers with students' typical conceptual development in the context of particular problem types would help teachers interpret ideas they were seeing in the moment and decide how to respond instructionally (Franke et al., 1998).

Already a distinction is clear between CGI and the video club work: CGI orients teachers toward a specific set of problem-solving strategies children may use, whereas the video club work aims to draw teachers' attention to students' mathematical thinking more generally. In fact, van Es (2011) explicitly noted that her "framework does not specify the kinds of student thinking that groups of teachers may analyze, such as student errors or solution strategies" (p. 149). The projects' schemes for evaluating teachers' progress reflect these differences. Nonetheless, there are core similarities, as we now discuss.

In assessing CGI participants' levels of instructional sophistication, researchers focused on the extent to which teachers' classrooms were organized around problem solving and the ways in which teachers worked with students' ideas (Fennema et al., 1996; Franke et al., 2001). Again, stance and specificity can be seen in the articulated levels. For instance, Level 1 teachers provided students with strategies to use and monitored how well students used the provided strategies. As one Level 1 teacher described, she typically "demonstrated the steps of a computational algorithm on the board and then asked her students to practice performing the steps. If they had difficulty, she would demonstrate the steps again using slightly different numbers" (Fennema et al., p. 415). In this account, the teacher must have attended to student activity in order to evaluate how they were doing with the steps but did not delve into students' answers in detail—an evaluative stance toward student reasoning attended to at a low level of detail. Level 2 teachers provided some opportunities for students to engage in independent problem solving and "could often repeat… what the children said" (Fennema et al., p. 416) but did not probe students for further or more complete explanations—attending to students' thinking at a higher level of detail than Level 1 teachers did, but still relatively superficially.

Level 3 teachers questioned students and "described student thinking in general terms; when they were specific about a child's thinking they would talk about the types of strategies the child used" (Franke et al., p. 668), reflecting attention to students' thinking at a still higher level of detail and interpretation centered on strategies. Level 4 teachers "articulated step-by-step descriptions of the strategies a student used to solve a given problem, and they routinely talked about the understanding underlying the student's solution" (Franke et al., p. 668). In other words, Level 4 teachers both described what students did in detail and interpreted what students' actions implied about their conceptual understandings—attending to ideas at a high level of detail and taking a strongly interpretivist stance toward their understandings. Researchers also highlighted a finer-grained distinction within Level 4 teachers. Level 4-A teachers took knowledge about groups of children into account in planning instruction, whereas Level 4-B teachers (the highest in the CGI scheme) "had more detailed knowledge of each child's thinking than Level 4-A teachers and seemed always to be aware of the impact that instruction would have on each individual" (Fennema et al., p. 421).

In these descriptions of levels, we can see progress from a limited and evaluative focus on student activity to a specific and interpretive analysis of students' strategies. Moreover, specificity has two components: being aware of details of individual students' thinking, as in other descriptions of responsive teaching, and recognizing in students' solutions the specific strategies in the research-based CGI frameworks. This second component of specificity separates the CGI scheme from the video club scheme discussed above. Note that a later iteration of CGI-like professional development focused even more on this second sense of specificity, coding whether teachers "identified the mathematically significant details" (Jacobs et al., 2010, p. 179) like how students counted or decomposed numbers and whether teachers' interpretations were consistent with research-based frameworks.

**Section summary**

In summary, the literature on teacher noticing/attention/responsiveness primarily defines progress as movement from (a) evaluating students' ideas, focusing on surface features to determine alignment with expected responses and making follow-up moves to push students toward particular ideas, to (b) interpreting students' meaning, focusing on the details of students' ideas and making follow-up moves to elicit more information from students. Most of the literature discusses students' ideas in general, with the exception of CGI's specific focus on students' problem-solving strategies.

Note that in this literature, considerations of disciplinary productivity and authenticity (Engle & Conant, 2002) – the extent to which a teacher notices and responds to more or less productive disciplinary *practices* with respect to student thinking – are not incorporated into conceptualizations of progress in responsive teaching. Certainly, the CGI scheme captures the level of detail at which teachers attend to students' strategy use, but it does not capture different disciplinary practices that could be involved in a student's use of a particular strategy – for example, whether the student is applying a memorized strategy, reasoning about the fit of a previously-encountered strategy, or inventing a new strategy.

In other pockets of literature, by contrast, considerations of disciplinary productivity and authenticity play a more central role in researchers' discussions of teachers' attention and instructional moves. We now turn to those accounts.

### Disciplinary considerations in rich descriptions of reform-oriented instruction

While research focusing on teachers' attention and responsiveness to student thinking has not teased apart different disciplinary aspects of student thinking and how teachers attend and respond to the distinctions between those aspects, other pockets of literature on science and mathematics teaching directly attend to these considerations.

**Seeing multiple possibilities in student thinking**

One such pocket of literature comes primarily from researcher-practitioners describing tensions in their own teaching around various ways in which students' ideas intersect with disciplinary ideas and practices (e.g., Ball, 1993; Chazan & Ball, 1999; Hammer, 1997). For example, Hammer reflected on the tensions he experienced when teaching an inquiry-oriented high school physics class. In one classroom episode, a student had engaged in productive debate to reach a canonically correct explanation for an electrostatic phenomenon. Next, she wanted to test her explanation empirically. Hammer worried about letting the student pursue the empirical route, which was a disciplinarily authentic path but was likely to lead to "noisy" experimental data which might weaken or even break the student's conviction in her conceptual explanation. Reflecting on another moment in the same class, Hammer discussed how he was primarily focused on the substance of students' causal explanations about the electrostatic phenomenon and on maintaining a climate of scientific argumentation rather than checking whether students' ideas were canonically correct. However, Hammer's goals did include helping students understand central scientific concepts, and in deciding what to do in the moment he often felt tensions among his goals of supporting students' engagement in causal explanation-building, coherence-seeking, empirical investigation, and coming to understand canonically correct concepts and explanations.

Ball (1993) also reported feeling such tensions when engaged in the sort of improvisational teaching involving "twin imperatives of responsiveness and responsibility" (p. 374) – grounding instruction in students' ideas, while helping them learn important disciplinary ideas and practices. In her well-known account of "Sean numbers," Ball described being unsure of whether to foreground students learning the canonical definition of even and odd numbers (and associated disciplinary practices of treating mathematical definitions in certain ways) or students engaging in their own definition formation and argumentation around those definitions. In this episode, Ball decided to prioritize the latter.

What we want to highlight in the above accounts is teachers' attention to multiple disciplinary aspects of student thinking – a single student idea could be (and often was) considered through varied lenses, with numerous possibilities for moving forward. Teachers' responsiveness involved tension-filled decisions about which disciplinary practices might be most productive to pursue.

**Promoting particular disciplinary norms**

A second, overlapping pocket of literature comes from extended descriptions of classroom interactions, presented (in part) to illustrate teachers' and students' roles in establishing and maintaining disciplinarily rich learning environments (Chazan & Schnepp, 2002; Cobb, Yackel, and Wood, 1989; Herrenkohl, Palinscar, DeWater, & Kawasaki, 1999; Hutchison & Hammer, 2010; Lampert, 1990; van Zee & Minstrell, 1997). For example, Cobb et al. investigated how teacher-student and student-student interactions led to norms around the social practices of mathematics in the classroom. Their descriptions highlighted that teachers' responses to students are instrumental in highlighting and valuing specific aspects of mathematical thinking, such as the adequacy and clarity of a mathematical explanation as opposed to a (mere) solution. Teachers' responses to students – what they question and what they leave unchallenged, ideas and moves they praise and those they don't – help students distinguish between different kinds of mathematical reasoning and engage in specific disciplinary practices.

Other descriptions of reform-oriented instruction also highlight the teacher's role in noticing, highlighting, and encouraging authentic disciplinary practices. For instance, in one of three ways in which Schnepp worked with his calculus students (Chazan & Schnepp, 2002), he aimed to engage students in authentic exploration and solution sharing/negotiation around novel problems:

> I respond to student questions with requests for more information about their thinking or simply ask them what other group members expressed in response to the same question. The explicit message I want to send is that they must do the thinking and that articulate, reasoned arguments are the goal (p. 174).

In an example from a physics course for preservice elementary teachers, Hutchison and Hammer (2010) described how Hutchinson attended and responded to cues indicating whether his students were framing their activity as making sense of natural phenomena or something else such as "playing school." For instance, he noticed and encouraged when students were using everyday language that made sense to them (an indicator of sense-making) as opposed to unfamiliar formal vocabulary.

In all of these extended examples of reform-oriented instruction, the authors emphasize how teachers preferentially attended and responded to certain *aspects* of the substance of student thinking in order to promote particular disciplinary practices and norms. Our point so far is that, while mathematics and science education researchers have discussed this kind of preferential attention and response, notions of what constitutes progress toward more responsive teaching as laid out in the teacher noticing/attention/responsiveness literature have not incorporated these kinds of disciplinary nuances. In this paper, we aim to start a discussion about how such disciplinary nuances could be incorporated into conceptualizations of progress in responsive teaching. To do so, we rely on a case study of a teacher's shift from preferentially attending and responding to students identifying *causal factors* to preferentially attending and responding to students telling *causal stories*.

## Data and Methods

### Background, subject selection, and episode selection

The data in this paper come from a professional development project aimed at helping fourth through eighth grade teachers promote inquiry teaching and learning in their science classrooms. Teachers voluntarily applied and often continued in the project for multiple years. As part of the project, teachers attended a two-week summer workshop in which they engaged in their own minimally-guided inquiry, discussed classroom video of students discussing scientific phenomena, and collaborated on lesson planning, assessment strategies, and other issues related to engaging their own students in inquiry. During the school year, teachers worked one-on-one with members of our research team to facilitate scientific inquiry in their classrooms and attended bimonthly small-group meetings with other participating teachers and members of the research team.

During the first summer of the project, our research team identified Mr. S as someone who seemed to fluctuate between engagement in deeper vs. more superficial coherence-seeking and explanation-building. In particular, we noticed that he often engaged in identifying *factors* causally relevant to the phenomenon under investigation, only sometimes building causal *stories* (mechanistic explanations, see Russ, Scherr, Hammer, & Mikeska (2008)) out of the factors. During the second summer workshop, by contrast, the research team believed that his participation was more often centered around creating and debating causal stories. We wondered if this apparent shift influenced what he focused on in *students'* explanations and took a closer look at science discussions Mr. S facilitated during the two academic years following these summers[1].

This article compares two classroom episodes from year 1 (April 2010) and year 2 (March 2011) of Mr. S's participation in the project. Specific features of this pair of episodes made them well-suited to exploring possible shifts in the aspects of student thinking to which Mr. S (and other science teachers) might attend. In many respects, the episodes are similar: They feature the same teacher teaching the "same" lesson to classes of seventh-graders at a Title I middle school[2]. In both episodes, Mr. S posed the same starter question: If you're walking at a steady pace while holding keys in your outstretched arm, and you want to drop the keys into a container sitting on the floor, should you release the keys before reaching the container, from directly over the container, or after passing the container? And in both episodes, he facilitated an extended classroom discussion about this question. Students posed sensible reasons for each option, and Mr. S elicited and entertained a range of possible answers. Yet, as we argue below, Mr. S foregrounded different aspects of students' disciplinary reasoning in year 1 vs. year 2.

**Analytical approach**

Our first step was to transcribe the two videotaped episodes, each approximately fifteen minutes in length. (Full episode transcripts are included as appendices.) The transcript captures pauses and emphases in participants' speech, drawing on transcriptional notations from Sacks, Schegloff, and Jefferson (1974): pauses in speech are indicated by long dashes (representing a beat) or (pause) (indicating a longer pause). Moments when a participant cuts himself off are represented by short dashes, and moments when a participant extends a word are represented by repeated colons in the middle of the word (e.g., "thi:::nk"). Emphases in speech are indicated by either underlined or capitalized words, with the latter representing increased volume specifically. Combinations of emphases and colons reflect pitch change in the course of a word, and () indicates that the speech could not be deciphered.

We then compared Mr. S's attention and responses to students' ideas across the two episodes. In doing so, we focused on exchanges in which students in both episodes expressed similar ideas. We drew on three kinds of evidence to unpack what Mr. S was preferentially attending and responding to during these exchanges:

- *How Mr. S revoiced students' ideas (Forman & Ansell, 2002; O'Connor & Michaels, 1993)*. Speech emphases (with respect to tone or volume) in his revoicings were used to infer what he was primarily attending to in a given moment (e.g., "Maybe GRA::vity. GRA::vity" [April 2010] vs. "Gravity's pulling it down" [March 2011]).
- *How and when Mr. S pressed on students' ideas (Brodie, 2011)*. The particular thrust of a question was used to infer what specifically Mr. S wanted students to flesh out (e.g., "So you're saying some kind of forward motion based on what?" [April 2010] vs. "Why will the keys go fast too?" [March 2011]).
- *When Mr. S made verbal and nonverbal bids to close the conversation (Jordan & Henderson, 1995; Schegloff & Sacks, 1999; Stivers & Sidnell, 2005)*. Accepting students' ideas as sufficiently articulated demonstrated what he found satisfactory (e.g., moving to another student's idea after a student identified wind as influential vs. after a student explained *why* wind was influential).

Evidence from other data sources—debrief conversations with Mr. S, Mr. S's recollections offered during small-group meetings, and stimulated recall/reflection interviews (Lyle, 2003)— were used to confirm or disconfirm our interpretations of what Mr. S was foregrounding in each episode. These data streams were consistent with our interpretations.

We also used these data streams to explore possible influences on Mr. S's attention and responsiveness in each episode. Specifically, we reviewed video in which Mr. S reflected on his interactions with students during the aforementioned episodes and noted points he raised that may have influenced the nature of his attention, such as how the discussion fit into his plan for the day. We also searched our data stores for examples of Mr. S posing his own explanations for scientific phenomena (which occurred most often during the summer workshops) and describing what he looked for in others' explanations. However, our analysis of Mr. S's own inquiry was not comprehensive, and conclusions based on these data are speculative.

### Analysis, Part 1: A shift in Mr. S's attention

In this section, we present and analyze several interactions between Mr. S and his students to illustrate our analysis of the two focal classroom episodes, one from Mr. S's class in April 2010 (episode 1) and the second from Mr. S's class in March 2011 (episode 2). We make the case that, in the two episodes, Mr. S foregrounded different aspects of students' scientific reasoning in his attention and responses. In the first episode, Mr. S's responses foregrounded causal *factors* responsible for the motion predicted by students. In the second episode, Mr. S's responses foregrounded causal *stories* explaining the motion predicted by students. We analyze how this shift in Mr. S's attention and responses to student thinking might have come about in a subsequent section.

**Mr. S foregrounded causal factors in episode 1**

In the first key drop episode (April 2010), Mr. S primarily attended and responded to a particular form of scientific knowledge with respect to students' ideas, namely the causal factors (such as force-like entities) responsible for the motion they predicted. Throughout the episode, Mr. S pressed students to articulate the factors underlying their ideas and accepted their explanations once this occurred. Here, we provide two examples to illustrate Mr. S's foregrounding of causal factors and also briefly point to supporting evidence from other exchanges throughout the episode.

The following exchange occurred at the beginning of the conversation when Mr. S asked students to consider why you might drop the keys right over the container. A student, Jack (all names are pseudonyms), mentioned the weight of the keys:

1. Mr. S: Uh, Jack?
2. Jack: The weight of the keys.
3. Mr. S: ((faces board, writes)) The weight. What's, say a little bit more about the weight. What is it about the weight?
4. Jack: The weight – the weight weighs ().
5. Mr. S: Hold on, not everybody's listening to your, to, to Jack right now. (pause) The weight of the keys will do what?
6. Jack: Wouldn't it make it go down because it's heavier?
7. Mr. S: So, so something having to do with the, the weight of the keys because it's heavy. What force will cause it to go straight down? What force will cause it to go straight down? ((Suri raises hand)) Suri?
8. Suri: Gravity.
9. Mr. S: ((faces board, writes)) Maybe GRA::vity. GRA::vity.

In the exchange above, Mr. S initially pressed Jack to say more about the weight in a relatively open-ended way: "What is it about the weight?" (line 3) "The weight of the keys will do what?"

(line 5) A critical shift, however, occurred in line 7. Mr. S revoiced Jack's idea about the keys being heavy, then made a strong press for identification of the factor or force related to Jack's idea: "What force will cause it to go straight down?" When Suri stated gravity, Mr. S excitedly accepted the response (line 9, "Maybe GRA::vity. GRA::vity."). At this point, Mr. S effectively closed this portion of conversation through a combination of physically turning away from students to write on the board and ceasing to ask questions that pursued this line of reasoning further.

      Further evidence of Mr. S foregrounding causal factors in relation to students' ideas comes from an exchange around another idea that came up in both key drop episodes, that the speed of the runner would make the keys move forward. In episode 1, a student, Diane, related this scenario to what would happen if you were to jump out of a racecar:

10. Diane: No, no, I'm not for that one, I feel like I would go before.
11. Mr. S: Before. Why before? You're for the first option.
12. Diane: Yeah.
13. Mr. S: Why before, Diane?
14. Diane: Because I thi:::nk that – well, let me try to give you an example, li:::::ke ((loudspeaker interruption)) I think, like, when you're racing? Like, you're in a racecar? And then, you know, let's say you have to () on fire or something? So when you're trying to land on the grass – because you're not going to get there right when you're at the grass or else you're gonna- because the car's fast, and you're going fast too. You gonna, like, get on the mud or something, so you're going to have to go before, so you know, you could, you know what I mean?
15. Mr. S: So what do you mean is that there's some kind of forward motion?
16. Diane: Yeah.
17. Mr. S: ((faces board, writes)) Okay. So you're saying some kind of forward motion based on what?
18. Diane: On the speed of the person who ().
19. Mr. S: So based on sp::eed, right?

In line 13, Mr. S's question ("Why before?") elicited a detailed story from Diane, but one with the causation only hinted at (e.g., line 14, "the car's fast, and you're going fast too"). His follow-ups, however, did not acknowledge Diane's specific details (racecar, grass, etc.) nor try to draw out a fuller causal story. Rather, he clarified that she was predicting the keys would go forward (line 15) and pressed Diane to identify the causal factor responsible for the motion (line 17). His verbal emphasis on Diane's identification of "sp::eed" as the relevant causal factor (line 19), followed by his moving on to other ideas, suggest that he was satisfied with this level of explanation.

      There were several other points throughout the 15-minute episode in which Mr. S pressed for or focused on specific factors or forces underlying the motion students described. For example, when a student Katherine talked about the keys going backward if you're going fast, Mr. S asked, "If I'm going fast, why would that cause the keys to go backwards? What, what force, what would cause the keys to go back?" His reframing of the question from why the keys would go backward to what force would cause the keys to go backward, and his subsequent summary that Katherine "said something about the wind," emphasized causal factors. Later in the conversation, Mr. S also asked students to argue against each other's ideas and focused on the factors they were comparing, making statements such as, "you're saying, Diane, that the angle is the key ((faces board, writes)), not the air push back" without further exploration or

justification.

**Mr. S foregrounded causal stories in episode 2**

When Mr. S explored the same key-drop question with another group of students in March of 2011, during his second year in the project, he primarily attended and responded to a different form of scientific knowledge with respect to students' ideas – their causal stories of how and why their predictions would occur. This involved articulation of relevant causal factors but also increased emphasis on and pursuit of explanations connecting the factors to the motion predicted. As before, we provide two illustrative examples.

As the discussion started, many students thought you should drop the keys from directly over the container. They offered multiple kinds of non-mechanistic explanations, including restatements of their conclusions (e.g., "Because if we drop it before or after the container, it won't get in the container") and appeals to the skill of the person dropping the keys (e.g., "Some people have bad aim, so they can't even aim towards the trash can"). Among these explanations was the following causal story from a student, Cooper:

20. Mr. S: Um, Cooper?
21. Cooper: Um, above?
22. Mr. S: Above.
23. Cooper: Because like the gravity, like, when you put it up, it goes down.
24. Drake: It's heavy...
25. Mr. S: Cooper said that because it's heavy, what happens, Cooper, I have to, I have to drop it-
26. Cooper: No, gravity puts, like, pulls it down.
27. Mr. S: So, because gravity's pulling it <u>down</u>.

Here, Cooper offered both a causal factor and how it works: gravity is the factor that works by pulling things down (line 26). In revoicing Cooper's response, Mr. S verbally emphasized what gravity does (line 27, "gravity's pulling it <u>down</u>") instead of emphasizing the causal factor only as occurred with "<u>GRA</u>::vity" in episode 1.

As students continued to offer different kinds of explanations, Mr. S returned to Cooper's explanation, emphasizing its causal nature: "So now let's, we want to get back to – why, why above? Cooper, you had some explanation why, what's the reason for it?" Mr. S recapped Cooper's response for a third time as he asked students for other reasons why you should drop the keys over the container: "Are there any other reasons why I should drop it above the container, other than Cooper said, the gravity's gonna pull it down. Why else might I drop it above the container?" This repeated emphasis on "why" and the causal story (what gravity does) instead of just the causal factor (gravity) suggest that Mr. S was interested in students providing mechanistic explanations.

This push beyond causal factors became most apparent in an exchange with Chavez about the speed of the runner making the keys move forward:

28. Chavez: If you do it before, it'll go directly in? But if you do it like, like-
29. Mr. S: <u>Why</u> do we have to do it before again?
30. Chavez: Because it'll go, like, IN, like the keys will go in the trash can or the thing will go in the trash can.
31. Mr. S: What will <u>cause</u> it to go in the trash can if we drop it before as opposed to over, because earlier you said over?
32. Chavez: Like, like, like, like, like the speed of the keys also I guess coming off.

33. Mr. S: The speed of the- so the keys have speed?
34. Chavez: Because you're walking, no, because like you're walking? (pause) And like, and like since you're walking fast, like, I guess the keys will also go fast too?
35. Mr. S: The keys will go fast too?...
36. Chavez: (pause) Yeah.
37. Mr. S: Why will the keys go fast too?
38. Chavez: I don't know!
39. Mr. S: I released the keys, wouldn't the keys just be there?

Recall how the exchange between Mr. S and Diane went in episode 1 when the idea of speed came up. Mr. S emphasized Diane's identification of speed and moved on to another student. Here, Mr. S's response differs notably, despite the parallels between Diane's idea that "the car's fast, and you're going fast too" (April 2010) and Chavez's idea that "since you're walking fast... the keys will also go fast too" (March 2011). First, in episode 2, Mr. S did not simply accept the idea of speed; he started to repeat it (line 33) but then reflected the idea back to Chavez with a questioning intonation (lines 33, 35). Second, Mr. S pushed Chavez to flesh out the story by asking, "Why will the keys go fast too?" (line 37). This question, followed by Mr. S's counterpoint that the keys might "just be there" once they're released (line 39), indicates that Mr. S was interested in more than the identification of speed as a causal factor. He also wanted Chavez to tell a causal story for how the keys would still have speed after they'd been released. In this case, it was Chavez's lack of response that closed this portion of the conversation, leaving Mr. S's question unanswered.

In summary, we have illustrated a difference between Mr. S's attention and responses to student thinking in episode 1 vs. episode 2. Though he attended to both causal factors and causal stories in both episodes, he foregrounded factors in episode 1 and stories in episode 2. We return to why this shift matters in the discussion section; here, we discuss whether this shift is already represented in the responsive teaching literature.

**Can published "responsive teaching" schemes capture the shift in Mr. S's attention?**

Schemes for characterizing responsive teaching point to a number of similarities across the two episodes. For instance, in both cases Mr. S was engaged in interpreting individual students' ideas; he neither directed the conversation toward the correct answer nor listened passively. And in both episodes, his most common follow-up moves were what Brodie (2011) highlighted as reform-type: *maintaining* a focus on students' ideas and/or *pressing* for more information. So, neither the *stance* dimension nor the types of follow-up moves that Mr. S deployed adequately capture the shift.

By contrast, one could argue that the shift in foregrounding from causal factors to causal stories aligns with a favorable shift along the *level of detail* dimension. Intuitively, foregrounding causal stories necessitates attention to details of students' explanations in a way not required by attention to causal factors. However, it's not clear how to identify the *level of detail* in cases where Mr. S attends to a whole story but then foregrounds only the relevant causal factor, or when he draws out details from a student offering a tangential argument (not incorporating causal factors or stories). Furthermore, even if Mr. S did attend to more detail in episode 2 than in episode 1, "attention to a greater level of detail" does not fully capture the nature of the difference between Mr. S's attention and responses in the two episodes. An important part of what differed across the episodes was the *sorts* of details Mr. S pressed *for*, namely details about causal factors vs. details about causal stories. ***In other words, there was a shift in which disciplinary practice Mr. S foregrounded in his attention and responses to students' ideas—*** *identifying causal factors vs. telling causal stories*—rather than simply in the level of detail to which he attended and responded to student thinking.

### Analysis, Part 2: What influenced the shift in Mr. S's attention?

As detailed in our discussion section below, our main point in this paper is to advocate for an additional aspect of progress in responsive teaching, one that captures shifts in the core disciplinary practices to which teachers attend and respond in students' reasoning. We used the case of Mr. S to illustrate what such a shift can look like. This argument does not rest on an explication of why such shifts in attention do or do not occur or what contributes to them. However, if our revised conceptualization of progress in responsiveness is to be useful for researchers and teacher educators, then it is worth exploring factors that may contribute to such progress and building causal stories for how such progress might happen.

In what follows, we present our analysis of factors contributing to Mr. S's shift between episode 1 and episode 2, even though our results are partial (more causal factors than causal stories!) and somewhat speculative. Drawing on video from various reflective conversations with Mr. S and his participation in inquiry during the summer workshops, we identified three factors that likely contributed to his shift. First, the nature of Mr. S's own explanations of scientific phenomena shifted between year 1 and year 2. Second, the key drop discussion played a different role in Mr. S's overall lesson plan in each case. Third, Mr. S structured the two discussions differently.

**The nature of Mr. S's own explanations shifted between year 1 and year 2**

During the first summer workshop, before Mr. S first asked his students the key drop question, Mr. S and his colleagues grappled with the key drop question themselves. Mr. S's explanations often centered on factors influencing the motion of the keys and whether these factors potentially interacted with each other. For instance, on the first day of the inquiry, Mr. S and other teachers discussed how water would fall from a crop plane on a windless day. Mr. S offered that the momentum of the plane, temperature, and air pressure would all matter in determining what would happen to the water, but with little detail about how or why. A bit later, the same small group discussed a similar but idealized scenario in which there was no air. Mr. S highlighted that "two of the main factors" would be the altitude and speed of the plane. During a whole-group discussion several days later, Mr. S stated, "I think there are many factors" and offered the following comparison:

> … if we increase the speed, or keep the speed constant at a certain level and increase the weight, at some point the impact of gravity on the weight of the object's going to be greater than the momentum causing the object to go forward.

Here, Mr. S identified several factors relevant to the object's motion: speed, weight, gravity, and momentum. He seemed to be specifically considering a situation in which speed was kept constant and weight was constantly increasing, and he thought there would be a point at which gravity (causing the object to move down) would overcome momentum (causing the object to go forward). This kind of tagging of the effects of different factors (e.g., momentum causes the object to go forward) and comparison of factors typified his style of extended explanation in the first summer when he went beyond naming causal factors.

In short, during the first summer workshop, Mr. S's participation was consistent with the claim that he considered identifying causal factors and their relative importance to be sufficient contributions to inquiry discussions. While we cannot fully substantiate this claim, we did not observe Mr. S struggling to flesh out causal stories explaining how or why the factors resulted in certain kinds of motion, nor indications that he considered articulating such causal stories to be important. We therefore speculate that Mr. S's attention to the causal factors underlying/within his students' ideas during episode 1 was influenced by his sense of what constituted a satisfactory scientific explanation at the time.

During the second summer workshop (occurring after episode 1 but before episode 2), Mr. S perceived that the professional development team (including the authors of this paper) emphasized the idea of mechanism more directly than we had the previous summer. This perception influenced and/or was influenced by Mr. S's own inquiry and his sense of causal explanation. For instance, one of the physical science inquiries during the second summer involved the pendulum pictured in Figure 1, with a peg approximately halfway between the attachment point and the ball at the end of the string when the ball is just hanging. Teachers were asked to predict how high the pendulum ball would swing when released from a height lower than the peg. Mr. S predicted that the pendulum would swing higher than the point of release:

> The reason I think it's gonna go higher… is because – when you lower it from, when you drop it... that second pin, somehow the – momentum that was in the first part of the string is going to be transferred to that second part of the string, and it will have more momentum – after it hits that pin, causing that part to go higher… because it accelerates the speed of the string, and by accelerating it, it'll cause the, uh – the ball to go higher.

As in the key drop inquiry during the first summer workshop, momentum was a key consideration for Mr. S. But here, his explanation involved more of a story of how momentum influenced the motion of the pendulum ball: once the string hits the pin (peg), the momentum formerly in the top of the string would transfer to the bottom half of the string, accelerating that part of the string and thereby causing the bottom of the string—and hence the ball—to go higher. Although this explanation is incorrect and much causality remains to be fleshed out, Mr. S started to unpack more of the causal connections between the identified causal factor (momentum) and the expected outcome (the pendulum ball swinging higher).

Evidence that the role of mechanism in causal explanations became more salient to Mr. S in year 2 comes from his participation in bimonthly small-group meetings, most notably in an emergent debate with another teacher, Ms. R. This debate occurred four months before episode 2 of Mr. S's own teaching (analyzed above). At the meeting, teachers were looking at Ms. R's students' written work about sinking and floating. Ms. R presented a rubric she invented to assess students' explanations. Mr. S questioned why Ms. R considered "causal story" and "mechanism" to be distinct:

40. Mr. S: So the, so [the student] is saying that it's sinking because water's going through the holes, that's not a causal story?
41. Ms. R: That's her, I took it as that's her mechanism of what the holes are doing.
42. Mr. S: So, but how is it not a causal story? It's an explanation of how it takes place, how it floats, how it sinks, right?

Later, Ms. R gave a clearer sense of what she meant by "mechanism," and Mr. S again related this to his sense of "causal story":

43. Ms. R: Mechanism is how is it working, what's causing it to, like the bicycle moving.
44. Mr. S: See, what I think is that your, from what you just said, mechanism is what we've been talking about as a causal story.

From these interactions, we see evidence that by the time episode 2 occurred, Mr. S's sense of "causal story" included explaining how something takes place, the underlying mechanism.

Further evidence for the salience of causal stories in Mr. S's thinking after year 1 comes from a stimulated recall interview. Two years after episode 2, Jen and Mr. S watched video of the episode, and Mr. S reflected on his aims for the lesson:
45. Mr. S: I was basically trying to get them to, to, to, to weigh in all the potential factors and also to, um, to come up with some kind of causal story as to how and where the, the item should be dropped. What are those factors, and uh, trying to get them to think more deeply about the movement of the, of the keys as related to the container.
46. Jen: Okay, so like the factors are part of an explanation-
47. Mr. S: Right.
48. Jen: And the causal story is relating the factors to –
49. Mr. S: The causal story, the causal story would, would utilize those various factors in its explanation as to how, how the keys would fall.

Here, Mr. S indicated that causal stories incorporate the sorts of causal factors upon which he focused in year 1: "the causal story would... utilize those various factors" (line 49). By this account, identification of relevant causal factors is not the endpoint but rather part of the process of telling causal stories about the movement of the keys. To be clear, we do not take this reflection as evidence that Mr. S was consciously thinking about causal factors and causal stories at the time of the second key drop episode. We do take it as evidence, however, that the notion of causal stories (as distinct from causal factors) was a "sticky" idea for him.

In this subsection, we have argued that after year 1, the notion of "causal stories" became more salient to Mr. S, a salience that manifested itself across multiple contexts. We speculate that this change in his own thinking about causal explanations contributed to his shift from foregrounding causal factors in episode 1 to foregrounding causal stories in episode 2.

**The key drop discussion played a different role in the overall lesson in year 1 vs. year 2**

Ayush's field notes from April 2010 indicate that, in the first key drop episode, the plan for the day was to "draw three [plausible] trajectories of the falling keys and take kids' reasoning again on each trajectory… Then to ask them to think carefully about how they want to test their idea and what the test outcomes could tell." So, in the lesson plan, a primary role of the key drop discussion was to set up ideas (predictions) for students to test experimentally. When designing and conducting such experiments, identifying causal factors, particularly ones that can be manipulated experimentally, is crucial. For instance, if students think speed matters, they could drop keys while walking slower vs. walking faster. If they think gravity is a factor and that it depends on the object's weight, they could work with heavier vs. lighter objects.

Mr. S showed evidence of thinking along these lines during episode 1. While making a bid to transition to the experimental design part of class, he explicitly asked students about relevant factors:
50. Mr. S: So what's a common theme- what's a common factor that we need to look at?
51. Student: Can we test it?
52. Mr. S: Yeah, but as we test it, what is something we need to look at? What's a common factor we need to look at?
53. Student: Speed.

Note that the first factor a student stated, in line 53, was the one Mr. S extracted from Diane's complex story about jumping from a car. Based on this evidence and on the overall lesson plan, we find it likely that Mr. S's press for causal factors during episode 1 was influenced by his sense that such factors would become useful in designing experiments later in the period.

By contrast, in the second key drop episode, Mr. S did not plan for students to test their ideas. In fact, he pushed back when a student suggested doing so:
54. Drake: Can we try an experiment?

    55. Mr. S: Well not ((holds hand toward Drake)), maybe not, maybe-

This episode took place on what Mr. S called "inquiry-based Monday," in which the whole period was devoted to *discussing* a scientific phenomenon. During a small-group meeting after episode 2, Mr. S reflected on this change from year 1 in his classroom practices:

> Before it was like part of a lesson, so I wanted to make sure that, that I, that I had, like, an inquiry part of the lesson, and then I would get to the exploration part of the lesson? As opposed to let the inquiry sit- that, that's the key difference. This year, the inquiry is, is kind of sitting alone by itself, connected to what happens during the week, but not – not so integrated to it that, that the inquiry can't take its own, go in its own direction, you know?... I think when we, when we made a space for the other possible causes, causal stories, uh, the kids have been – so far, you know, they've been, they've been coming up with them, you know?

Here, Mr. S stated that inquiry-based Monday allowed the inquiry to takes it own direction. This more open version of inquiry created a space for other "possible causes, causal stories" that may not have had space to emerge the previous year. Without an experiment to get to, Mr. S was freer to follow students' ideas for an extended period of time to flesh out their causal stories, and he had no obvious incentive (e.g., the need to identify manipulable variables) to focus on causal factors.

**The discussions were structured differently**

    In episode 1, Mr. S engaged students in a whole-class discussion, which he noted was similar to what he experienced as a participant in the key drop inquiry during the first summer workshop. (We should note that teachers in the summer workshop spent much of their time in small-group discussion and activities as well; apparently the whole-group discussions were particularly salient for Mr. S.) He also recorded students' ideas on the board. Though he referred to the board in clarifying his understanding of students' ideas and later solicited counterarguments to specific ideas, the act of recording may have reinforced his focus on causal factors (and the associated predicted trajectories), as he could jot them down quickly and populate the board.

    In the second key drop episode, Mr. S did not take notes and used a "fishbowl" discussion structure. Seven students sat in an inner circle, the "fishbowl," and discussed the key drop scenario while the rest of the class sat in an outer circle to listen and reflect on the inner circle's discussion. (Students rotated into and out of the fishbowl over the course of the period.) At a small-group meeting, Mr. S indicated that he's "able to listen more clearly to what kids are saying" when there are fewer students. By this account, the "fishbowl" structure of episode 2 afforded Mr. S the opportunity to delve more deeply into fewer students' thinking at a time, as compared to the whole-class format of episode 1.

## Discussion and Conclusion

**A summary of the case study**

    In this paper, we documented a shift in Mr. S's attention and responses to the substance of student thinking in two consecutive years of teaching a similar lesson. The lesson focused on explaining the trajectory of keys dropped by someone in motion, so that the keys have some initial horizontal speed. In the key drop inquiry in year 1, Mr. S foregrounded students' identification of the causal factors (in this case, force-like entities) responsible for the motion they predicted, such as gravity or forward speed or momentum. In year 2, by contrast, Mr. S foregrounded students' articulation of causal stories—the beginnings of mechanistic accounts of why the keys would follow predicted trajectories, e.g., gravity pulling the keys down, or a

mechanism (solicited but never supplied) for how the keys have forward speed after the walking person drops them. We argued that previous conceptualizations of progress in responsive teaching in the literature do not capture this kind of shift in attention to different disciplinary practices in which students may engage (identifying causal factors vs. telling causal stories). We then explored likely influences on Mr. S's foregrounding in each case, highlighting a shift in (i) his own scientific explanations over the two years, toward greater emphasis on causal stories; (ii) the role of the key drop discussion, from setting up an experiment to standing alone as discussion-based inquiry for its own sake; and (iii) the structure of the discussion, from whole-class discussion in conjunction with writing on the board to a fishbowl (reducing the number of students to whom Mr. S needed to attend at a given time).

In the following sections, we make two arguments. First, drawing on literature about the nature of science in the context of science education, we make the case that in the context of the key drop discussion, attending and responding to causal stories is more sophisticated than attending and responding to causal factors. We propose that this kind of shift from foregrounding causal factors in students' explanations to foregrounding causal stories represents progress in responsive teaching – progress that considers the nature of the disciplinary practices teachers notice and respond to in students' ideas. Second, we explore what would count as evidence for such a shift in responsive teaching to reflect stable progress vs. a locally-triggered change (unlikely to persist). Within each section, we address the implications of these arguments for continuing professional development and research on responsive teaching.

**Why care about the shift in attention from causal factors to causal stories?**

In science education, scholars have emphasized the importance of students constructing causal explanations of natural phenomena as a way of building understanding and engaging in scientific practice (e.g., Chinn & Malhotra, 2002; Hammer & van Zee, 2006; Russ, Scherr, Hammer, & Mikeska, 2008; Sandoval, 2003; Windschitl, Thompson, Braaten, & Stroupe, 2012). For instance, Chinn and Malhotra (2002) draw on work from the psychology, sociology, philosophy, and history of science to argue that one aspect of authentic inquiry is "the development of theoretical mechanisms with entities that are not directly observable" (p. 186). Sandoval's (2003) analysis of causal coherence in students' scientific explanations also focused on causal mechanisms, how students chain causes and effects to create coherent explanations. In creating and discussing causal stories of how or why something happened, students engage in a practice that is arguably at the core of science.

Generally speaking, fleshing out causal stories (when feasible) is a more sophisticated form of scientific explanation than simply identifying relevant causal entities or factors (Russ et al., 2008; Windschitl et al., 2012). For instance, Russ et al. (2008) developed a framework for analyzing students' mechanistic reasoning, adapted from work in philosophy science that was based on studies of explanations in the scientific literature. Russ et al.'s framework highlights the importance of identifying entities and properties and actions of entities relevant to the target phenomenon; however, more sophisticated mechanistic reasoning also involves creating a coherent chain of how these activities or properties bring about the target phenomenon. In other words, identifying causal factors contributes to but is less sophisticated than telling causal stories, which requires consideration of how the factors behave and interact with each other over time.

That said, there are certainly situations in which foregrounding the identification of relevant causal factors is appropriate, like when engaging in experimental design (e.g., Ford, 2005; Toth, Klahr, & Chen, 2000). Identification of causal factors provides useful insights about phenomena and predictive power with respect to similar phenomena, and is a publishable finding in various scientific disciplines, such as ecology and epidemiology[3]. We do not mean to argue that pursuing causal stories is uniformly more desirable or sophisticated than pursuing causal factors.

In the context of a key drop discussion, however, pursuing causal stories *is* more scientifically authentic, for reasons that Russ et al. (2008) and others articulate: such explanations provide more predictive and explanatory power than mere identification of factors. For instance, identifying gravity as a factor does not distinguish whether gravity always pulls things straight down or sometimes pulls objects along a curved downward path, a possibility we have heard in many key drop discussions. By contrast, a more defined causal story that gravity pulls the keys straight downward leads to the conclusion that either the keys fall straight down or that the overall causal story for the motion of the keys must involve other factors. For this reason, even in a case such as episode 1 when the key drop discussion leads quickly to an experiment, telling causal stories is scientifically productive because such stories increase the interpretability of the experimental results (see Schauble, Glaser, Raghavan, & Reiner (1991) for a discussion of relations between the sophistication of students' causal models and their processes of experimentation).

Philosophy of science aside, does it really matter *in the classroom* if teachers foreground attention to causal stories vs. causal factors in their interactions with students' ideas? As Cobb et al. (1989) point out, what teachers attend and respond to in the classroom can influence what norms take hold. One classroom period is too short a period to see such evolution occur, but even over the course of episode 1, we saw glimmers of this process. Shortly after Mr. S recapped Diane's idea as having to do with speed, Ayush (who was visiting the classroom that day) asked, "Folks, did you hear that reasoning?" A student responded, "Yes, it's based on speed," suggesting that Mr. S's foregrounding of causal factors may have been taken up by at least one student in that moment. If we want students to learn various scientific practices such as identifying causal factors and telling causal stories, and if we want students to distinguish between them and make reasoned choices about which practices to enact in which contexts, then it is critical to explore how teachers are promoting and distinguishing between these disciplinary practices in their attention and responses to students' ideas. Moreover, it is critical to frame this as part of the work of responsive teaching – of hearing and building on possibilities in students' ideas in ways that acknowledge connections between their thinking and the discipline, including disciplinary practices. Future research could continue to unpack these nuances of teachers' responsive practices, as well as connections between teachers' foregroundings and students' understandings of the local activities in which they are involved and what they see or come to see as authentic disciplinary practice.

A skeptic could agree with this argument in the abstract but worry that helping science teachers distinguish causal factors from causal stories is too nitpicky and difficult, especially given the need to help teachers take an interpretive stance toward student ideas *at all*. We agree that helping teachers to move beyond only listening for correctness and to attend more deeply to students' meanings and perspectives are top priorities. However, the responsive teaching community has developed a variety of approaches for fostering an interpretive stance toward student thinking (e.g., Sherin and colleagues' video clubs; discourses and tools for eliciting and analyzing student thinking (Thompson, Windschitl, & Braaten, 2013)), and we anticipate that the time is ripe for considering how to help teachers notice, interpret, and foster the seeds of core disciplinary practices in students' ideas. For instance, part of the discussion around student thinking could involve identifying various disciplinary aspects that participants (including professional developers) note with respect to students' ideas and considering what to foreground in a given moment and why. Making this kind of thinking explicit could promote teachers' awareness of perhaps tacit foregroundings in their interactions with students and may facilitate progress in responsive teaching in two senses: in foregrounding more sophisticated disciplinary practices with respect to student thinking (as in the case of Mr. S), and in shifting among foregrounding different disciplinary aspects of students' ideas more responsively during classroom inquiry.

**Stable progress vs. context-dependent shifts in responsive teaching**

In this last subsection of our discussion, we turn to another issue in the responsive teaching literature onto which our analysis of Mr. S can shed light. Even when a later episode of teaching is judged as "more responsive" than an earlier episode, the question remains: is it reflective of stable *progress* toward more responsive teaching, or is the difference in responsiveness merely a result of a difference in local conditions such as subject matter, student population, etc.? We first consider how the responsive teaching literature speaks to this issue. Then, we use Mr. S to illustrate what we think should count as evidence for such a shift being categorized as stable progress vs. local and contextual, ultimately arguing for the need to couple cognitive and social considerations in doing so.

**Stability *and* variability in responsive teaching.** Studies of teacher noticing and responsiveness illustrate tensions with respect to attributing stability to observed shifts in teacher practice. For instance, Brodie (2011) observed four mathematics teachers' classroom practice every day for one week, classifying and counting the types of follow-up moves teachers used in interaction with students' ideas during whole-class discussions. After an intervention, during which teachers had an opportunity to jointly plan lessons aimed at engaging learners' mathematical thinking, Brodie observed their practice for another week. Though Brodie saw overall shifts in teachers' responses toward more reform-oriented moves, she also expected and recognized hybridity in their take-up. Her analysis acknowledged several specific examples of variability in their responsiveness, including variability that she associated with the nature of the tasks teachers posed (different tasks elicited different student contributions and called for different teacher responses) and variability in one teacher's responses to students' errors: "In a number of cases in week 2, Mr. Peters reverted to his week 1 style in response to the additional errors" (Brodie, p. 184). Furthermore, we note that it is unclear whether the overall shifts observed would persist outside the context of Brodie's study; teachers' jointly planned lessons and knowledge of the study's focus on student thinking may have made them more conscious than usual of eliciting and encouraging student thinking during the post-intervention observations. We simply cannot tell if the shifts noted in the context of the study indicate more stable changes in the teachers' practices or not.

Similarly, van Es and Sherin (2008) examined teachers' discourse in response to video clips of classroom teaching (as described previously) across ten meetings over the course of a year. van Es and Sherin broke each meeting into segments and coded for each teacher's primary focus along each of their dimensions in a given segment. By combining the segments and comparing the percentages of foci across meetings, van Es and Sherin noted that "teachers' analyses of video shifted in terms of who and what they found noteworthy, how they analyzed these interactions, and their level of specificity" (p. 253). Interviews around video segments corroborated these results, and similar results were not seen for a control group of teachers not involved in the video clubs.

However, van Es and Sherin (2008) noted different pathways by which video club teachers accomplished these shifts, one of which was "cyclical" (p. 258) – shifting back and forth between focusing broadly on a number of considerations and focusing narrowly on student thinking. The teacher exemplifying this pathway tended to focus broadly on clips from other teachers' classrooms and more narrowly on the student thinking evident in clips from his own classroom, suggesting that specific video clips and their origins may themselves represent importantly different contexts with respect to attention to student thinking. This pathway highlights how shifts toward a greater focus on student thinking are not straightforward nor necessarily stable, even within the video club context. Similar dynamics played out when the researchers extended their study to the classroom setting (Sherin & van Es, 2009; van Es & Sherin, 2010); by analyzing classroom observations early in the year and late in the year, they saw significant shifts in the extent to which teachers created space for and followed up on students' ideas, but they also noted disconfirming evidence of these practices. For instance, they described the case of Linda as follows:

> Late in the year, we observed several instances in which Linda investigated the meaning of her students' ideas and methods. Yet within the same lessons, there were also moments in which Linda did not attend to students' comments and made no attempt to reason about what her students were thinking. Thus, although she was certainly capable of noticing student thinking and reasoning in depth, this was not a consistent aspect of her practice (Sherin & van Es, p. 31).

This description cuts to the heart of the questions we are raising about stability. Although Linda demonstrated shifts toward more responsive teaching between early and late classroom observations, she still exhibited a good deal of variability *within* her late classroom observations. How should we as a community of researchers think about whether Linda has made stable progress in responsive teaching?

We bring up these issues not to critique Brodie or van Es and Sherin's work, but rather to illustrate the difficulties in teasing apart stable progress from context-dependent shifts and to highlight how examining trends across a number of observations does not automatically solve the problem. For instance, in both Brodie's and van Es & Sherin's studies, all of the observations took place in the context of invested researchers visiting teachers' classrooms. In this context, it is possible that teachers may have tailored their instruction toward what they had learned about the researchers' aims *without* making changes to their day-to-day instructional practice. Additionally, the studies above (and others, see Lau, 2010; Levin, 2008; Thompson, Windschitl, & Braaten, 2013) demonstrate relevant variability of attention and response associated with specific local conditions, such as the nature of the instructional task given to students. These possible contextual influences are critical to consider when trying to determine the nature of observed shifts in responsive teaching. Recognizing that our study runs into many of the same issues (e.g., observations by invested researchers, relevant contextual differences in how the discussions are situated, etc.), we now explore other ways of trying to distinguish stable progress from context-dependent shifts.

**Other sources of evidence for addressing stability vs. context-dependency.** We ground this discussion in our attempt to answer the following question: Was the change we documented in Mr. S between year 1 and year 2 a stable shift, likely to persist, or was it brought about by a difference in local conditions between the two episodes? While not conclusive, our data is suggestive of both. In laying out the evidence for both conclusions, our goal is not to reach a final answer about Mr. S, but rather to illustrate what kinds of evidence researchers can gather and analyze in trying to address the stability issue.

One argument for the stability of the shift centers on Mr. S's views about the role of mechanism in scientific explanations. As discussed above, Mr. S showed evidence of an epistemological shift toward ascribing a more central role to mechanism in year 2. An additional point we make here is that evidence of this shift in epistemology manifested itself in several contexts, including his own participation in inquiry in the summer workshops, his interactions with other teachers during project meetings, his informal conversations with researchers visiting his classroom, formal stimulated recall interviews with us, and his interactions with students (or so we argued). To the extent that this epistemological shift helped to drive Mr. S's increased attention and responsiveness to the seeds of causal storytelling in his students' ideas, the robustness of this epistemological shift across contexts and time can be imagined to contribute to the stability of the shift in responsive teaching.

One could still make the counterargument that all of the contexts in which we observed Mr. S professing and enacting a changed epistemology were associated with the professional development project. However, there were instances in small group meetings in which Mr. S reported on class discussions he facilitated on his own and described the mechanistic details of students' ideas. While the reporting occurred in the project context, facilitation of the discussion and attention to mechanism occurred when project staff were not present – a further indication of stability.

The use of structures like inquiry-based Monday and the fishbowl format (as opposed to whole-class discussion) in year 2 illustrates further subtleties that arise when trying to distinguish stable progress from context-dependent shifts in responsive teaching. At first glance, the fishbowl format seems to be a local contextual factor whose absence would dampen or even eliminate Mr. S's "progress." However, the fishbowl format was not a random fluctuation in Mr. S's classroom practice. From his conversations with Ayush and with other teachers, we know that his experience facilitating inquiry in year 1 alerted him to the challenges of responsive teaching and to his own strengths and weaknesses. In year 2, he actively worked to structure discussions to enable his pressing deeper into students' ideas; to this end (as described above), he noted that inquiry-based Monday allowed him to flesh out students' ideas without immediately worrying about tying them to canonical content, and the fishbowl allowed him to focus on and pursue a few students' ideas at a time. It is this strategic, self-aware decision-making on Mr. S's part – rather than the specific structures *per se* – that counts as evidence of Mr. S's progress in responsive teaching.

Yet, in other ways, inquiry-based Mondays *were* a local fluctuation that were not sustained. Inquiry-based Mondays arose partly because of Mr. S's interactions with a special education co-teacher who was in his room during year 2. The co-teacher was skeptical of inquiry but was willing to accept a mix of traditional and inquiry-based instruction, which may have influenced Mr. S's sequestration of inquiry onto a specific day. Additionally, other local conditions likely influenced Mr. S's instruction too – he was well-established and relatively comfortable at his school, and he worked regularly with Ayush through extended phone calls and visits on inquiry-based Mondays. Mr. S and Ayush typically co-planned the inquiry discussion, and during his visits, Ayush provided feedback to Mr. S – often honestly expressing his delight at how the discussion was going. Mr. S and Ayush also talked extensively about the specific causal stories that came up from students. All of these conditions may have helped to stabilize Mr. S's attention to student thinking and specifically to their causal stories. To decide to what extent Mr. S's responsiveness in year 2 was connected to these contextual factors, we would need to observe him in later years when some or all of these factors are absent.

Sadly, year 3 provided the opportunity to make such observations. First, the conditions at Mr. S's school changed. He only taught self-contained special education science classes (with many different co-teachers), and his administration made hands-on activities a priority, making it more difficult for him to integrate discussion. Second, with new teachers joining our professional development program, we were not able to allocate as much researcher time to veteran teachers; Ayush spent far fewer hours visiting Mr. S's classes and communicating with him outside of class. Conversations at teacher meetings indicated that Mr. S was no longer facilitating inquiry discussions in his class and was engaging in more traditional instruction, imbued with hands-on activities. He expressed disappointment and dissatisfaction with his teaching, both in teacher meetings and in private conversation with us, but felt he didn't have a choice. He continued to strongly advocate for inquiry – with attention to causal stories – in teacher meetings, and in subsequent summer workshops he continued to participate in inquiry in sophisticated ways. This evidence shows how important the local conditions were in producing the phenomenon we noticed in year 2.

**Blurring the distinction.** Our attempts to decide whether Mr. S's shift in responsiveness was stable illustrate the need for greater specificity about what "responsiveness" means. Defined as Mr. S's desire to be responsive and his epistemological views about the centrality of causal storytelling, Mr. S's increased "responsiveness" in year 2 showed indications of stability, as discussed above. Such affective and epistemological considerations would likely contribute to the stability of responsive behaviors by Mr. S when he teaches in a variety of supportive contexts. But with "responsiveness" defined in terms of observable classroom behaviors, our data indicate that the local context can have a strong influence in constructing the classroom phenomenon. In year 2, Mr. S was stable in his responsive teaching during inquiry-based Mondays, with this "context" and the associated stability emerging from his unfolding interactions with Ayush, his co-teacher, other teachers and researchers in our professional development project, his students, and others. On non-Mondays, he co-constructed other contexts with very different affordances, and in his classes in subsequent years, he co-constructed still other contexts with different sets of players. When various supports fell away in year 3, Mr. S no longer engaged in responsive teaching. However, it is also plausible that if the supports from year 2 were present in year 1, Mr. S still would not have demonstrated the kind of responsiveness – attention to causal stories in student reasoning – seen in year 2, given his epistemological orientation at the time. Our broader point here is that research on responsive teaching should aim to better understand how the social and the cognitive are coupled to produce and sustain (or disrupt) responsive teaching, with careful attention to the multiple meanings of "responsiveness" just discussed.

# INCORPORATING DISCIPLINARY PRACTICES 25# **References**

Ball, D. L. (1993). With an eye on the mathematical horizon: Dilemmas of teaching elementary school mathematics. *Elementary School Journal*, *93*(4), 373–397.

Bouyer, J., Coste, J., Shojaei, T., Pouly, J. L., Fernandez, H., Gerbaud, L., & Job-Spira, N. (2003). Risk factors for ectopic pregnancy: a comprehensive analysis based on a large case-control, population-based study in France. *American Journal of Epidemiology, 157*(3), 185-194.

Brodie, K. (2011). Working with learners' mathematical thinking: Towards a language of description for changing pedagogy. *Teaching and Teacher Education*, *27*, 174–186.

Carpenter, T. P., Fennema, E., & Franke, M. L. (1996). Cognitively Guided Instruction: A knowledge base for reform in primary mathematics instruction. *Elementary School Journal*, *97*(1), 3–20.

Carpenter, T. P., Fennema, E., Franke, M. L., Levi, L., & Empson, S. B. (2000). *Cognitively Guided Instruction: A research-based teacher professional development program for elementary school mathematics* (Research report No. 003). Madison, WI: National Center for Improving Student Learning and Achievement in Mathematics and Science.

Carpenter, T. P., Fennema, E., Peterson, P. L., Chiang, C.-P., & Loef, M. (1989). Using knowledge of children's mathematics thinking in classroom teaching: An experimental study. *American Educational Research Journal*, *26*(4), 499–531.

Chazan, D., & Ball, D. L. (1999). Beyond being told not to tell. *For the Learning of Mathematics*, *19*(2), 2–10.

Chazan, D., & Schnepp, M. (2002). Methods, goals, beliefs, commitments, and manner in teaching: Dialogue against a calculus backdrop. *Social Constructivist Teaching*, *9*, 171–195.

Chinn, C. A., Buckland, L. A., & Samarapungavan, A. (2011). Expanding the dimensions of epistemic cognition: Arguments from philosophy and psychology. *Educational Psychologist*, *46*(3), 141–167.

Cobb, P., Yackel, E., & Wood, T. (1989). Young children's emotional acts while engaged in mathematical problem solving. In D. B. McLeod & V. M. Adams (Eds.), *Affect and mathematical problem solving: a new perspective* (pp. 117–148). New York, NY: Springer-Verlag.

Coffey, J. E., Hammer, D., Levin, D. M., & Grant, T. (2011). The missing disciplinary substance of formative assessment. *Journal of Research in Science Teaching*, *48*(10), 1109–1136.

Crespo, S. (2000). Seeing more than right and wrong answers: Prospective teachers' interpretations of students' mathematical work. *Journal of Mathematics Teacher Education*, *3*, 155–181.

Davis, B. (1997). Listening for differences: An evolving conception of mathematics teaching. *Journal for Research in Mathematics Education*, *28*(3), 355–376.

Empson, S. B., & Jacobs, V. R. (2008). Learning to listen to children's mathematics. In D. Tirosh & T. Wood (Eds.), *Tools and processes in mathematics teacher education* (pp. 257–281). Rotterdam, NL: Sense Publishers.

Engle, R. A., & Conant, F. R. (2002). Guiding principles for fostering productive disciplinary engagement: Explaining an emergent argument in a community of learners classroom. *Cognition and Instruction*, *20*(4), 399–483.

INCORPORATING DISCIPLINARY PRACTICES	26...

INCORPORATING DISCIPLINARY PRACTICES 28Schegloff, E. A., & Sacks, H. (1999). Opening up closings. In A. Jaworski & N. Coupland (Eds.), *The discourse reader* (pp. 263–274). London, England: Routledge.
Sherin, M. G., & Han, S. Y. (2004). Teacher learning in the context of a video club. *Teaching and Teacher Education*, *20*, 163–183.
Sherin, M. G., Jacobs, V. R., & Philipp, R. A. (Eds.). (2011). *Mathematics teacher noticing: Seeing through teachers' eyes*. New York, NY: Routledge.
Sherin, M. G., & van Es, E. A. (2009). Effects of video club participation on teachers' professional vision. *Journal of Teacher Education*, *60*(1), 20–37.
Stivers, T., & Sidnell, J. (2005). Introduction: Multimodal interaction. *Semiotica*, *156*(1/4), 1–20.
Thompson, J., Windschitl, M., & Braaten, M. (2013). Developing a theory of ambitious early-career teacher practice. *American Educational Research Journal*, *50*(3), 574–615.
Toth, E. E., Klahr, D., & Chen, Z. (2000). Bridging research and practice: A cognitively based classroom intervention for teaching experimentation skills to elementary school children. *Cognition and Instruction*, *18*(4), 423–459.
van Es, E. A. (2011). A framework for learning to notice student thinking. In M. G. Sherin, V. R. Jacobs, & R. A. Philipp (Eds.), *Mathematics teacher noticing: Seeing through teachers' eyes* (pp. 134–151). New York, NY: Routledge.
van Es, E. A., & Sherin, M. G. (2008). Mathematics teachers' "learning to notice" in the context of a video club. *Teaching and Teacher Education*, *24*, 244–276.
van Es, E. A., & Sherin, M. G. (2010). The influence of video clubs on teachers' thinking and practice. *Journal of Mathematics Teacher Education*, *13*, 155–176.
van Zee, E., & Minstrell, J. (1997). Using questioning to guide student thinking. *Journal of the Learning Sciences*, *6*(2), 227–269.
Windschitl, M., Thompson, J., Braaten, M., & Stroupe, D. (2012). Proposing a core set of instructional practices and tools for teachers of science. *Science Education*, *96*, 878–903.

**Footnotes**

[1] To be clear, definitively establishing connections between Mr. S's evolving participation in inquiry and his classroom practice is not the aim of this paper, although plausible ties are explored.

[2] According to publicly available 2009-2010 demographic data, 65% of the students at the school identify as Hispanic, 30% as African American, and about 35% are classified as having limited English proficiency.

[3] See publications like "Changes in sub-alpine tree distribution in western North America: a review of climatic and other causal factors" (Rochefort, Little, Woodward, & Peterson, 1994) or "Risk factors for ectopic pregnancy: a comprehensive analysis based on a large case-control, population-based study in France" (Bouyer et al., 2003).